\documentclass[aps,prl,twocolumn,showpacs,superscriptaddress]{revtex4} 
\usepackage{graphicx}
\usepackage{amsmath}
\usepackage{color}
\newcommand{\YTO}{Yb$_2$Ti$_2$O$_7$}

\begin{document}

\title{Mapping the first-order magnetic transition in \YTO}
\author{E. Lhotel} 
\email[]{elsa.lhotel@neel.cnrs.fr}
\affiliation{Institut N\'eel, CNRS \& Universit\'e Joseph Fourier, BP 166, 38042 Grenoble Cedex 9, France}
\author{S. R. Giblin} 
\affiliation{School of Physics and Astronomy, Cardiff University, Cardiff, CF24 3AA, United Kingdom}
\author{M. R. Lees}
\affiliation{Department of Physics, University of Warwick, Coventry CV4 7AL, United Kingdom}
\author{G. Balakrishnan}
\affiliation{Department of Physics, University of Warwick, Coventry CV4 7AL, United Kingdom}
\author{L. J. Chang}
\affiliation{Department of Physics, National Cheng Kung University, Tainan 70101, Taiwan}
\author{Y. Yasui} 
\affiliation{Department of Physics, Meiji University, Kawasaki 214-8571, Japan}

\begin{abstract}

The very nature of the ground state of the pyrochlore compound \YTO~is much debated, as experimental results demonstrate evidence for both a disordered or a long-range ordered ground state. Indeed, the delicate balance of exchange interactions and anisotropy is believed to lead to competing states, such as a Quantum Spin Liquid state or a ferromagnetic state which may originate from an Anderson-Higgs transition. We present a detailed magnetization study demonstrating a first order ferromagnetic transition at 245 mK and 150 mK in a powder and a single crystal sample respectively. Its first-order character is preserved up to applied fields of $\sim$ 200 Oe. The transition stabilizes a ferromagnetic component and involves slow dynamics in the magnetization. Residual fluctuations are also evidenced, the presence of which might explain some of the discrepancies between previously published data for \YTO.

\end{abstract}
      
\maketitle

Magnetism affected by geometrical frustration is an active field due to the ability generate new and unusual magnetic phases ~\cite{Lacroix}. In this context, the pyrochlore oxide materials R$_2$M$_2$O$_7$ (R=rare earth, M=metal) form a very rich family in which a large diversity of new physics can be explored ~\cite{Gardner10}. Specifically, the rare-earth ions lie on the vertices of corner sharing tetrahedra, forming the highly frustrated pyrochlore lattice. Depending on the rare-earth element, the anisotropy of the spins as well as the exchange and dipolar interactions can be varied so that different model Hamiltonians can be studied within this structure. One of the most spectacular realizations is the spin-ice phase (R=Dy,Ho, M=Ti) \cite{Harris97, Ramirez99} in which the local spin arrangement obeys the ice-rule (two spins point into and two spins point out of every tetrahedron in the structure) and which possesses a macroscopically degenerate ground state. This state is induced by the strong uniaxial anisotropy along the local $<$111$>$ axes of the tetrahedra, combined with a resultant ferromagnetic interaction.   
With these ingredients and in the presence of strong transverse fluctuations, a new magnetic state is expected to be stabilized, the quantum spin ice (QSI) in which exotic excitations are predicted \cite{Onoda10, Savary12, Shannon12}. 

 \YTO~ has been proposed as a good candidate for stabilizing the QSI state \cite{Ross11a, Applegate12}. Indeed, the exchange in \YTO~ is highly anisotropic, with a strong ferromagnetic component akin to the Ising exchange of spin ice \cite{Ross11a, Cao09, Thompson11b}, despite an XY-like anisotropy perpendicular to the local  $<111>$ directions \cite{Hodges01, Malkin04}. At low temperature, using a model Hamiltonian with anisotropic exchange parameters deduced from experiments, a first-order phase transition towards a long-range ferromagnetic order is predicted \cite{Applegate12, Chang12, Savary13, Han13}. 

Experimentally, the existence of a long-range magnetic ordering in this compound is debated, suggesting a fragile ground state with respect to perturbations. In an early study, a peak was observed around 210 mK in the specific heat of a polycrystalline sample \cite{Blote69}. It was later shown to be associated with an abrupt slowing-down of the fluctuations in the low temperature phase thus suggesting a first-order transition \cite{Hodges02}. 

Below the transition, depending on the nature of the samples (single crystal or polycrystal) and the crystal growth conditions, different results have been obtained. Some neutron scattering measurements demonstrate ferromagnetic long-range order (LRO) \cite{Yasui03, Chang12} while others do not \cite{Gardner04, Bonville04, Ross09}. A discrepancy is also observed in muon spin relaxation measurements ($\mu$SR) where an anomaly at the transition is present \cite{Hodges02, Chang13} or not \cite{DOrtenzio13}.
In the meantime, it was shown that the peak in specific heat strongly depends on the samples \cite{Yaouanc11, Ross11b} so that the presence of a transition towards a long-range order might depend on the sample quality. 

It has been suggested that the specific heat anomaly, however, does not necessarily correspond to a magnetic ordering \cite{Ross11b, DOrtenzio13}.
It is therefore essential to probe another thermodynamic quantity, which should be more sensitive to the magnetic nature of the transition: the magnetization. 
In this letter, we show that the magnetization of  \YTO~presents a first-order transition in both a powder sample and a single crystal which was shown to develop additional magnetic intensity on structural peaks \cite{Chang12}. 
The first-order nature of the transition suggested in previous studies \cite{Hodges02, Chang12, DOrtenzio13} is proved by the existence of a small thermal hysteresis (of a few millikelvin width). The transition is accompanied by strong time-dependent effects. The magnetization value below the transition temperature is consistent with the stabilization of a ferromagnetic ordering with a reduced spontaneous moment, suggesting a strongly fluctuating spin component. Significantly the first-order behavior occurs below the peak in the specific heat where only a deviation in the susceptibility is observed. 

Three samples were measured: i) a compacted powder sample, ii) a crushed powder sample mixed with Cu grease to ensure a good thermal coupling, both obtained from the same synthesis; 
iii) a single crystal, with the magnetic field applied along the [100] and [110] axes. These samples are the same as those used in Ref. \onlinecite{Chang13} (and Ref. \onlinecite{Yasui03, Chang12} in the case of the single crystal). 
 Magnetization and a.c. susceptibility measurements were performed down to 70 mK on SQUID magnetometers equipped with dilution refrigerators developed at the Institut N\'eel \cite{Paulsen01}. 
At 70 mK, the magnetization of all the samples reaches about 1.7 $\mu_B$/Yb above 60 kOe \cite{SupMat}. 
The susceptibility follows a Curie-Weiss law down to 1.5 K \cite{SupMat}, consistent with previous results above 2 K \cite{Bramwell01, Hodges01, Yasui03, Thompson11a}. 
Below 1.5 K, the susceptibility still increases with the temperature decreasing but deviates from the Curie-Weiss law. Then at 250 mK for the powder (165 mK for the single crystal), the magnetization $M$ increases abruptly and  reaches a plateau at low temperature. This very sharp increase immediately suggests a magnetic transition, especially as it matches approximately with the anomaly in specific heat \cite{Chang13}.

In an ordered ferromagnetic state, a system has a spontaneous magnetization and the initial intrinsic susceptibility is expected to diverge. The measured value of the bulk susceptibility is then equal to the inverse of the demagnetizing factor $N$. 
For the four measurements performed (two powder samples and two orientations of a single crystal), the measured $M/H$ value (which can be considered as the susceptibility in such small fields) is roughly equal to the estimated value of $1/N$ \cite{SupMat}. This result shows that a spontaneous magnetization exists in \YTO~below the specific heat anomaly, thus implying the existence of an ordered ferromagnetic component in the system. It also suggests that the specific heat anomaly is associated with a magnetic transition. 
Analysis of the $M$ vs $H$ curves at low field gives an estimation of the ordered moment between 0.35 and 0.8 $\mu_B$/Yb, with a main component along the [100] direction, which implies that a fraction of the moment is still fluctuating in the ordered phase \cite{SupMat}. 
The obtained orientation might indicate that the magnetic structure is similar to the one determined in the related compound Yb$_2$Sn$_2$O$_7$ \cite{Yaouanc13}. 

\begin{figure}
\centerline{\includegraphics[width=7.0cm]{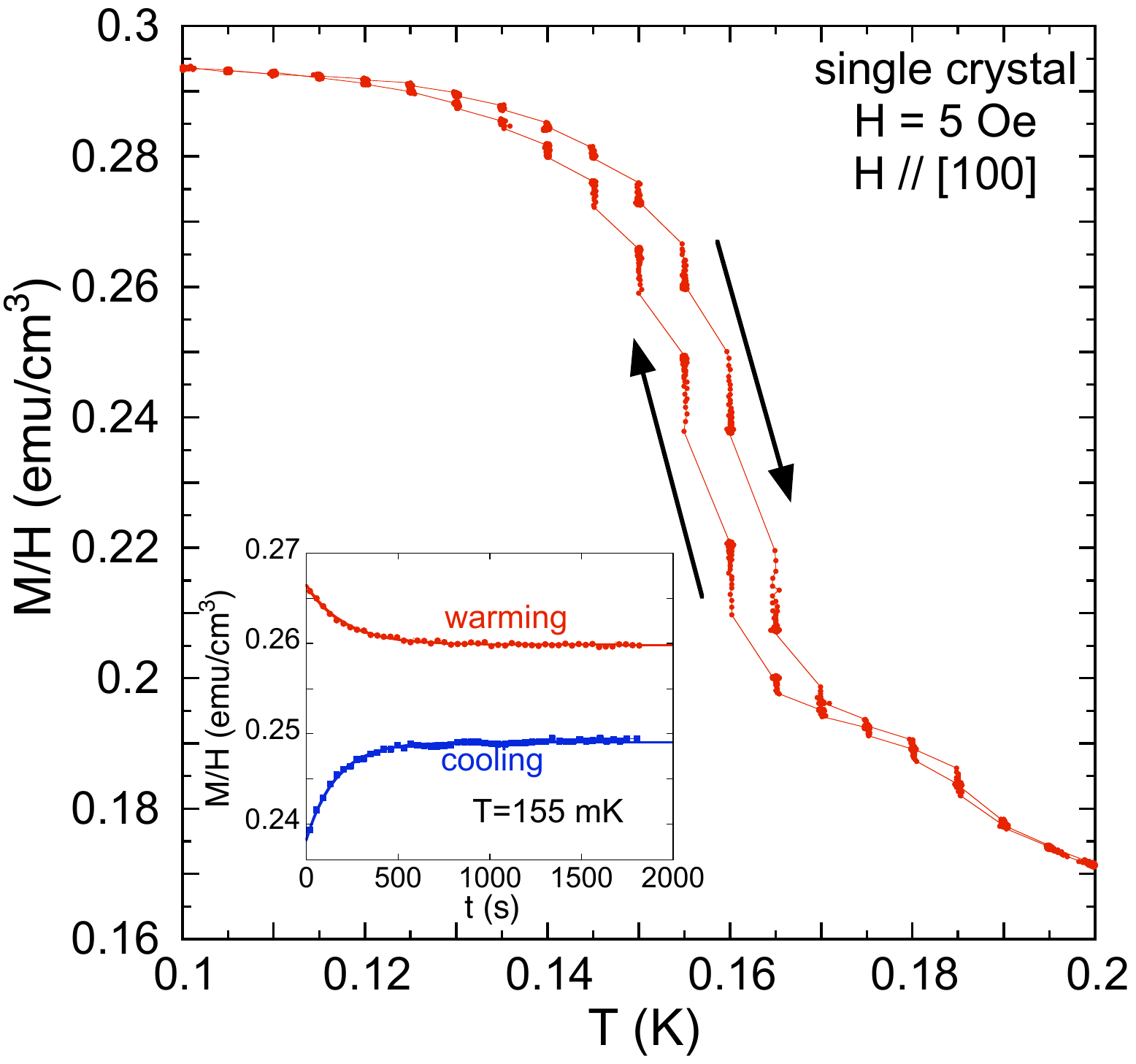}}
\caption{(color online) $M/H$ vs $T$ for the single crystal in an applied fied $H=5$ Oe parallel to the [100] axis at the proximity of the transition. 
The temperature was swept in steps of 5 mK and 100 extractions were made at each temperature ($\sim 30$ min at each temperature). 
Inset: isotherm as a function of time $t$ at $T=155$ mK when warming (red circles) and when cooling (blue squares). 
The lines are fitted to the exponential : $\frac{M}{H}(t)=\frac{M_{\rm eq}}{H} - \frac{\Delta M}{H} \exp(-t/\tau)$. 
When warming $\tau=207$ s, $\frac{M_{\rm eq}}{H}=0.260$ emu.cm$^{-3}$ and $\frac{\Delta M}{H}=-6.58 \times 10^{-3}$ emu.cm$^{-3}$. 
When cooling  $\tau=165$ s, $\frac{M_{\rm eq}}{H}=0.249$ emu.cm$^{-3}$ and $\frac{\Delta M}{H}=1.08 \times 10^{-2}$ emu.cm$^{-3}$.
}
\label{fig1}
\end{figure} 

A detailed study of the magnetization around the transition has been performed. To ensure accurate results, measurements had to be performed with well controlled temperature regulation and extremely slow cooling and warming rates. The protocol was the following: i) regulate at a given temperature. ii) take a large number of measurements (between 40 and 100) so that the magnetization reaches equilibrium at this temperature. iii) change temperature with a step of 5~mK, or 2 mK depending on the measurements. The temperature was ramped between 80 - 400 mK, cooling and warming the sample. The equivalent ramping rate is between 9 and 18 mK/h. The obtained magnetization as a function of temperature for the single crystal is shown in figure~\ref{fig1}: it can be seen that, at the transition, at a fixed temperature, a strong relaxation occurs. As shown in the inset of the figure where the magnetization is plotted as a function of time, at 155 mK, the equilibrium value is reached after times as long as 600 s. 

\begin{figure}
\centerline{\includegraphics[width=7.0cm]{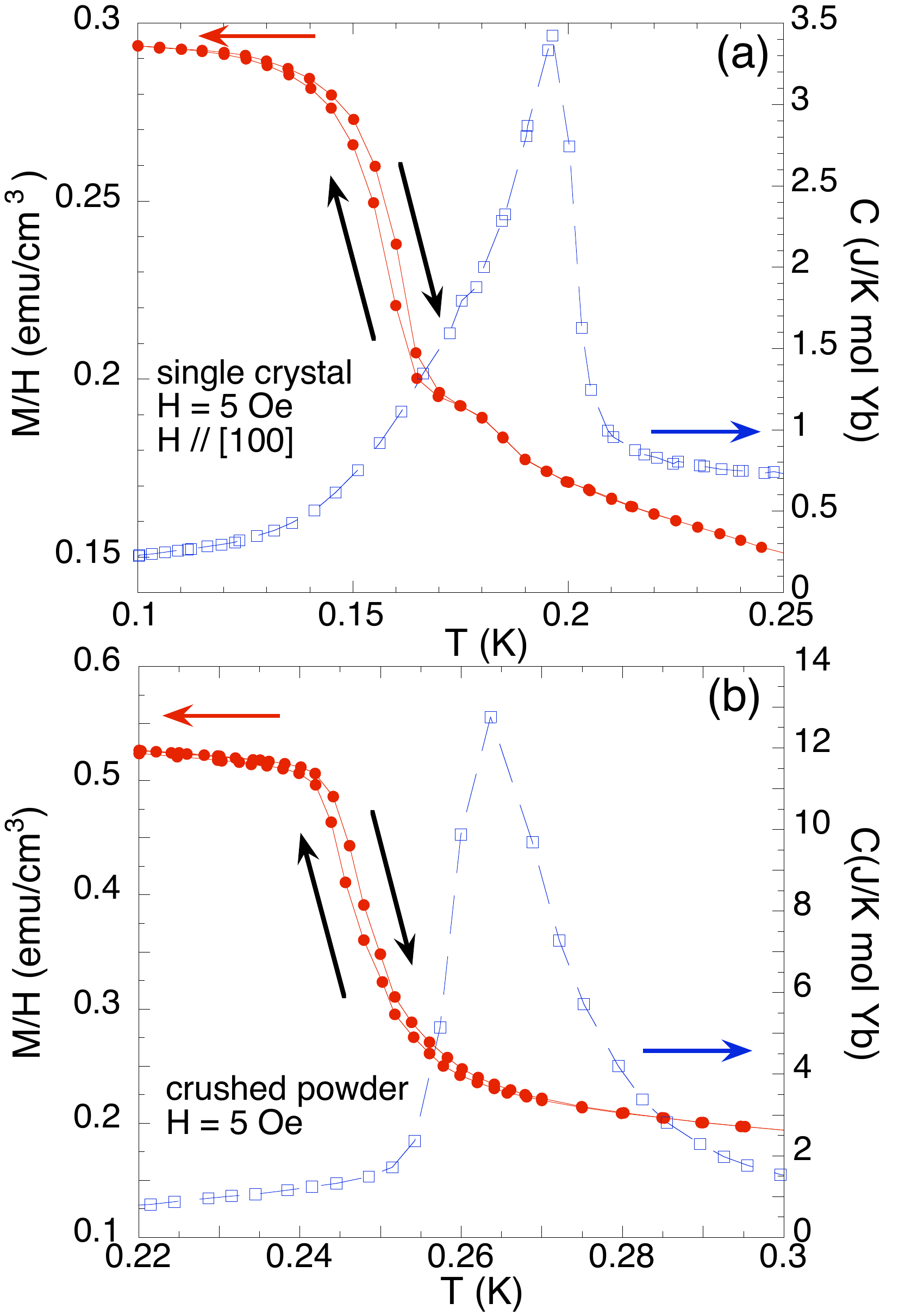}}
\caption{(color online)a. $M/H$ vs $T$  for the single crystal in an applied fied $H=5$ Oe parallel to the [100] axis, extracted from figure \ref{fig1} with only the equilibrium value of the magnetization plotted compared to the specific heat data. b. The equivalent data for the crushed powder.}
\label{fig2}
\end{figure}

Figure \ref{fig2}a shows the equilibrium values of the magnetization at the transition (obtained from figure \ref{fig1}) as a function of temperature for the single crystal. It can be seen that a small hysteresis is present (which is much narrower than for a fast temperature sweep), indicating a first-order like behavior. 
Also shown is the specific heat data on the same crystal. A subtle change of slope occurs in the magnetization at the peak in specific heat, while the first-order transition develops below this peak. 
The bump observed at $\approx$180 mK before the sharp increase is not present in the magnetization of the powder sample as shown in figure \ref{fig2}b and might be due to a sample inhomogeneity, a consequence of difficulties in sample preparation~\cite{Ross11b,Yaouanc11}.

From the magnetization, it appears, that the first-order transition occurs around 150 mK in this single crystal. The transition extends over about 20 mK and the hysteresis width is about 3 mK. For the powder sample, the transition occurs around 245 mK, but the width of the transition and of the hysteresis are similar. 
Zero Field Cooled - Field-Cooled (ZFC-FC) magnetization shows an irreversibility below the temperature of the transition \cite{SupMat}. In ordered materials, such irreversibility is often ascribed to domain freezing. The ZFC-FC irreversibility is strongly reduced when the applied field is increased and is suppressed at about 500~Oe. This suggests that small fields are enough to overcome the barriers involved in the freezing. 

\begin{figure}
\centerline{\includegraphics[width=7.0cm]{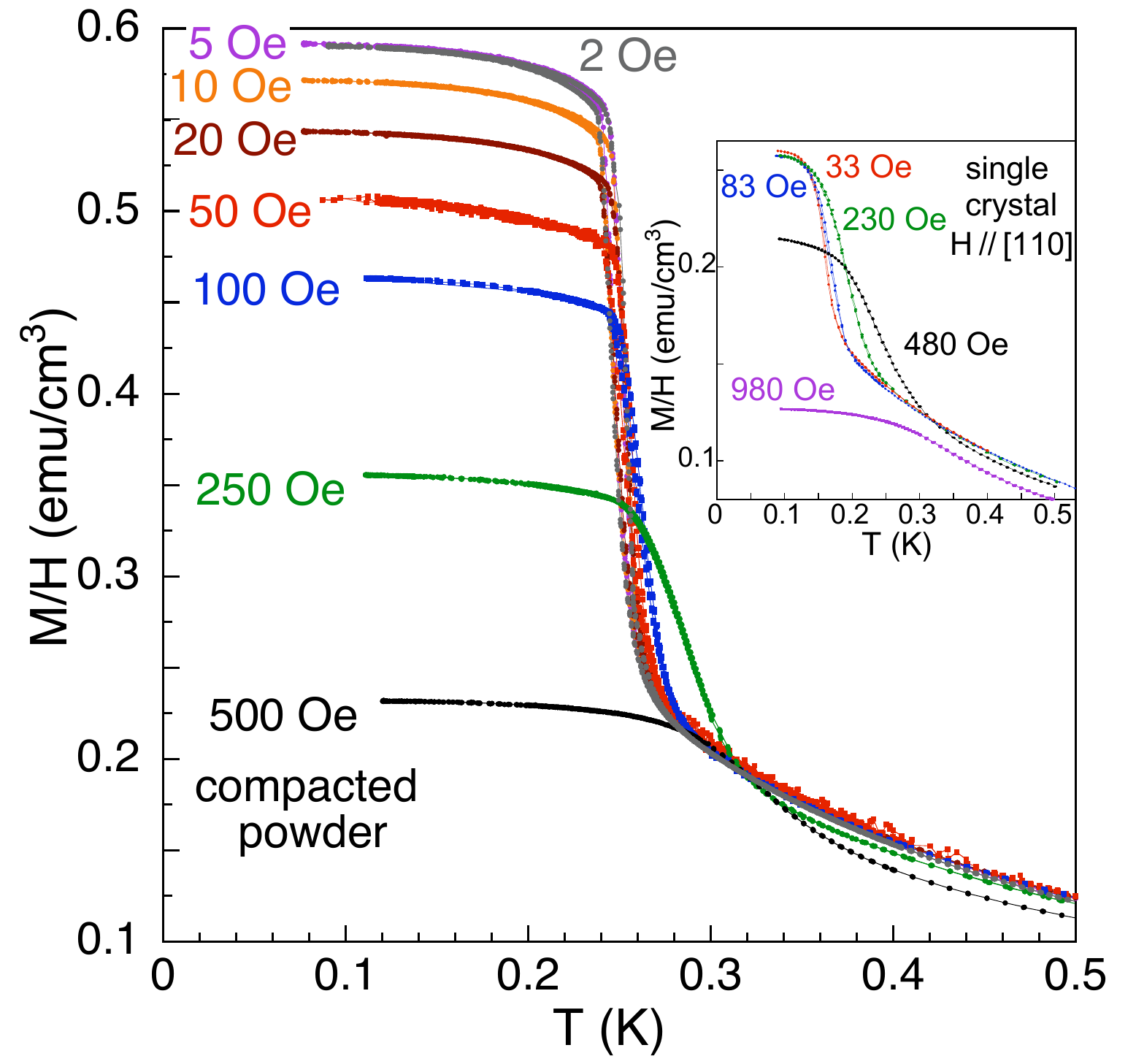}}
\caption{(color online) $M/H$ vs $T$  for the compacted powder for several fields between 2 and 500 Oe. Inset: $M/H$ vs $T$ for the single crystal for several fields between 30 and 1000 Oe, applied along the [110] direction. }
\label{fig3}
\end{figure}

The effect of the field on the transition has also been investigated.
For the powder samples, a field as small as 10~Oe is enough to reach the regime of nonlinear susceptibility, resulting in a smaller value of $M/H$ (see figure \ref{fig3}). However, up to at least 20 Oe, the transition temperature and the transition and hysteresis widths remain constant. At 50 Oe, the transition starts to become smoother and shifts to higher temperature, but the hysteresis persists. Above 250 Oe, the hysteresis is lost. The ``step" in the magnetization continues reducing in magnitude and broadening. 

The results are qualitatively similar in the single crystal (see inset of figure \ref{fig3}) and independent of the direction of the applied field. However, the transition in the single crystal appears to be less sensitive to magnetic field, except the bump at 180 mK which is suppressed in 10~Oe. The susceptibility remains linear and the transition is unchanged up to 30 Oe. Above 50 Oe, the transition broadens and starts to shift to higher temperature (which seems analogous to the reported behavior of the a.c. susceptibility in Yb$_2$Sn$_2$O$_7$ \cite{Dun13}). Above approximately 200 Oe, the hysteresis disappears (similar to the powder samples) and finally, the amplitude of the ``step" decreases significantly above 500~Oe. The main conclusions from this field induced behavior are: i) the hysteresis and so the first-order character is preserved up to an applied field of $\sim$ 200 Oe. ii) the increase of the applied field broadens and shifts the magnetization step to higher temperatures, as expected for ferromagnetic transitions. Above 200 Oe, the magnetization measurements alone cannot conclude whether the ``step" is a signature of a phase transition or rather of a crossover to a field polarized state. 

\begin{figure}
\centerline{\includegraphics[width=8.0cm]{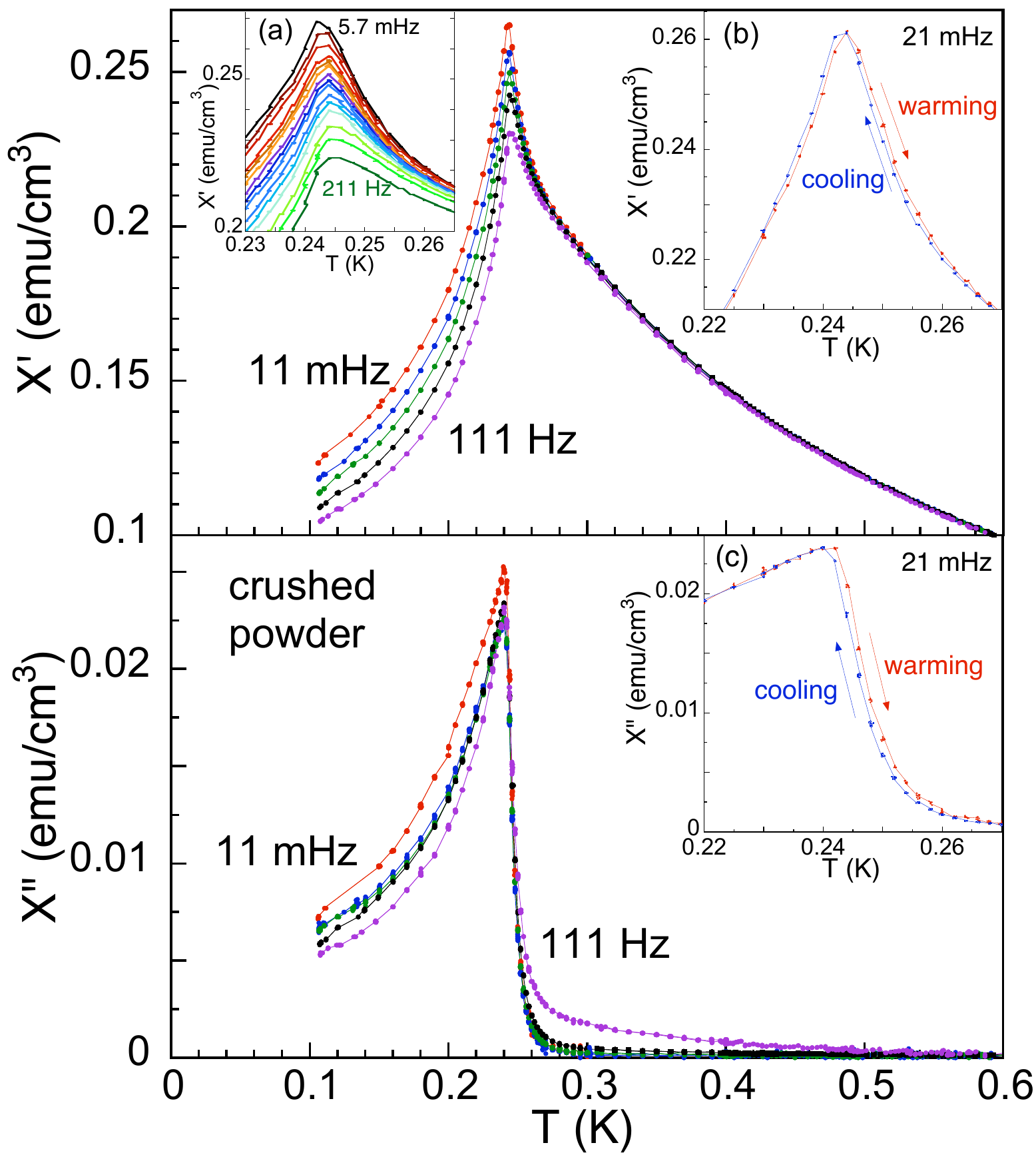}}
\caption{(color online) a.c. susceptibility, in phase $\chi'$ (top) and out-of-phase $\chi''$ (bottom), vs $T$ for the crushed powder sample. Data collected while cooling. The amplitude of the a.c. field is 1 Oe. (a) Magnification of the $\chi'$ peak for frequencies between 5.7 mHz and 211 Hz. (b) $\chi'$ and (c) $\chi''$ vs $T$ at the transition at 21 mHz for the cooling and warming ramps.}
\label{fig4}
\end{figure}

The dynamics at the transition have been studied via a.c. susceptibility, and are shown for the powder sample in figure \ref{fig4}. $\chi'$ shows a sharp peak, which is associated with the onset of the out-of-phase part $\chi''$ of the susceptibility.  In the measured frequency range (5.7 mHz - 2.11 kHz), no frequency dependence of the peak position was observed (in both the powder and crystal), but the amplitude of the peak increases a little when the frequency decreases (see figure \ref{fig4}a). A small hysteresis is observed in $\chi'$ and $\chi''$ (see figures \ref{fig4}b and  \ref{fig4}c), in the same temperature range as the magnetization. The whole characteristics of the a.c. susceptibility are consistent with the picture of a first-order magnetic transition. The $\chi''$ onset would then be the signature of the dissipation at the transition. The temperature of the $\chi'$ peak gives a transition temperature $T_c=243 \pm 1$ mK. These results are in strong contrast with a.c. measurements in Yb$_2$Sn$_2$O$_7$ \cite{Lago14}, where a glassy behavior was reported.  

Our results show an agreement between thermodynamic measurements, i.e. magnetization and specific heat, supporting a first-order transition in \YTO, at 245 mK and 150 mK for the powder and the single crystal respectively, and involving a LRO ferromagnetic component.
This is qualitatively consistent with neutron measurements performed in the single crystal \cite{Yasui03, Chang12}, although our results (transition temperature, hysteresis width, value of the ordered moment), do not match quantitatively. 

These features, a first-order transition resulting in a small ferromagnetic ordered magnetic moment suggesting a fluctuating component are reminiscent of other pyrochlore compounds which exhibit long-range order \cite{Gardner10}: Gd$_2$Sn$_2$O$_7$, the archetype of dipolar Heisenberg pyrochlore antiferromagnet \cite{Wills06, Bonville03} and Tb$_2$Sn$_2$O$_7$, an ordered spin-ice \cite{Mirebeau05}. In these compounds, persistent spin dynamics are observed but the ordered moment is much larger than in \YTO, most probably resulting in a more robust long-range ordering. 

The fragility of the ordered state in \YTO~results in a sample dependence of the transition. The reasons for this dependence are under debate, and are of great importance to understand the mechanisms of the ordering in \YTO. Questions immediately arise from the results reported here:  magnetization measurements definitely show that a first-order magnetic transition occurs in these samples around the vicinity of the peak in specific heat. Does suppression of the ordering temperature in the single crystal suggest that disorder induced by the growth process increases spin fluctuations? Do other samples which present a peak in specific heat but no evidence of long-range ordering in neutron scattering measurements \cite{Ross11b} also show such evidence in magnetization? The analysis of our single crystal results as a function of temperature may give a preliminary answer to this question: the presence of a reversible bump before the step in the magnetization, seems to indicate a partial ordering first, at the specific heat maximum, before the achievement of the transition at lower temperature. Thus it might be possible that, in other samples  \cite{DOrtenzio13, Ross11b} in which the microscopic probes (neutron scattering, $\mu$SR) do not detect LRO, such a partial ordering could occur at the specific heat peak but without ending in a transition at lower temperature. So that the existence of a peak in specific heat may not involve a long-range magnetic ordering.
If a LRO transition was also present in the magnetization in these samples, the presence of strong fluctuations might be the clue to explain why the thermodynamic probes (magnetization, specific heat) do show the evidence of a transition while the microscopic ones (neutron scattering, $\mu$SR) do not.  Further magnetization measurements would be needed in these samples to answer this question.

Finally, it is interesting to discuss the origin of the transition itself. Theoretical work predicts a first-order character of the transition \cite{Chang12, Applegate12, Savary13}, in agreement with the above results. However, the mechanism for the transition is still debated: a Higgs mechanism from a Coulomb phase \cite{Chang12}, or a confinement of the excitations from a thermal spin liquid state \cite{Savary13} for example have been proposed. It might be of great interest to consider theoretically the effects of the magnetic field on the transition in the different scenarios and to compare them with the dependence (temperature and order) on the magnetic field reported above.  

In conclusion, we have shown that in both powder and single crystal samples, a first-order magnetic transition occurs in \YTO. The transition width is about 20~mK while the hysteresis at the transition is about 3~mK. The value of the d.c. susceptibility below the transition indicates the existence of a spontaneous magnetization and so of a ferromagnetic component. The transition remains first-order up to about 200 Oe. In larger fields the magnetization anomaly softens and shifts to higher temperatures.

\acknowledgments 
C. Paulsen is warmly acknowledged for the use of his magnetometers. J. Debray is acknowledged for cutting and orienting the crystal. E.L. thanks S. Petit and B. Canals for fruitful discussions. S.R.G. greatfully acknowledges the support of the European Community Ñ Research Infrastructures under the FP7 Capacities Specific Programme, MICROKELVIN project number 228464.

\newpage
\onecolumngrid
\begin{center} {\bf \large Supplemental Material for \\
Mapping the first order magnetic transition in \YTO } \end{center}
\vspace{0.5cm}
\twocolumngrid

 \setcounter{figure}{0} 

\section{Samples and demagnetizing factors}
As stated in the main text, the reported magnetization and a.c. susceptiblity were measured on the same samples as Ref. \onlinecite{Chang13}: two powder samples, from the same synthesis, and a single crystal. The details of the single crystal growth are given in Ref. \onlinecite{Chang12}.  

The shape of these samples is irregular thus preventing an accurate correction for demagnetizing effects.
Using the formula provided for parallelepipeds \cite{Aharoni98}, the demagnetizing factors $N$ could be nevertheless estimated: \\
i) for the single crystal, $N$ is within the range  [2.7; 3.5] and [5.3; 4.7] (cgs units) for measurements along the [100] and [110] directions respectively. \\
ii) the compacted powder has an elongated shape, giving $N \approx 2$ (cgs units). \\
iii) for the crushed powder, the estimation is much harder, but $N$ should be larger.

\section{Magnetization vs Temperature below 4.2 K}
\begin{figure}[h]
\centerline{\includegraphics[width=7.0cm]{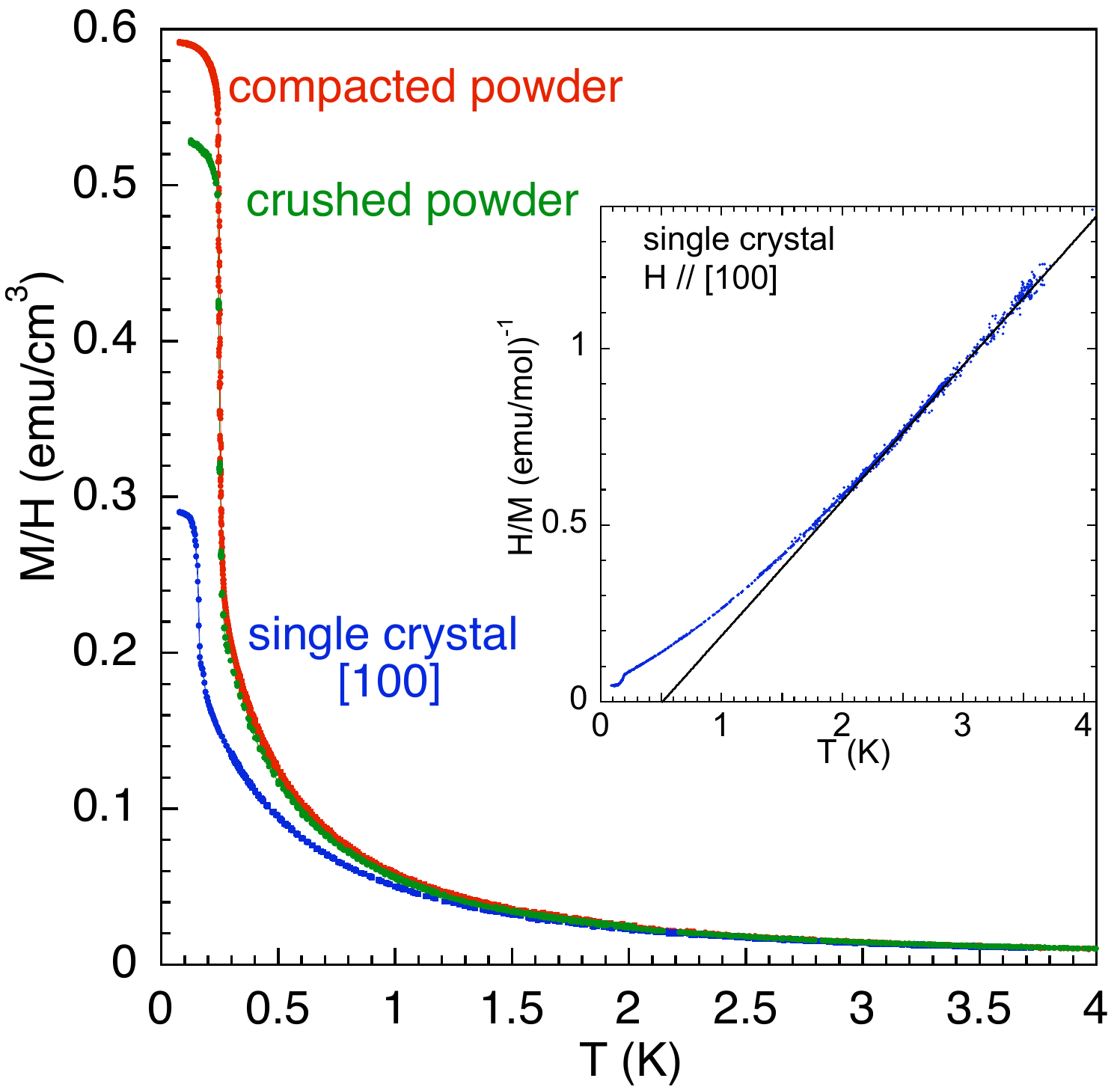}}
\caption{$M/H$ vs $T$ for the single crystal ($H$ // [100]) and for the two powder samples. The applied field was 5 Oe below 1 K and 100 Oe above (where the magnetization $M$ is linear in $H$ for such fields). Inset: $H/M$ vs $T$ for the single crystal. The line is a fit to a Curie-Weiss law $H/M$ (emu$^{-1}$.mol)$=-0.195 + 0.382~T$ for $T>1.5$ K. }
\label{fig_MT_HT}
\end{figure}

Figure \ref{fig_MT_HT} shows the magnetization measured in a small field as a function of temperature up to 4~K for all the samples. Above 1.5 K, the susceptibility follows a Curie-Weiss law (See inset of figure \ref{fig_MT_HT}), consistent with previous results above 2 K \cite{Bramwell01, Hodges01, Yasui03, Thompson11a}. Some differences are obtained between the field orientations. They are mainly due to different demagnetizing effects depending on the shape of the sample even if a small anisotropy is expected \cite{Hayre13}. 

Below 1.5 K, the susceptibility continues to increase with decreasing temperature but deviates from the Curie-Weiss law. Then at 250 mK for the powder (165 mK for the single crystal), the magnetization $M$ increases abruptly and  reaches a plateau at low temperature. 

It is worth noting that the magnitude of $M/H$ (which can be considered as the susceptibility in such a low field) on the plateau is larger for samples with the smaller demagnetizing factors, as expected, and is in the range of $1/N$. 
In particular, preliminary measurements were performed on a roughly parallelepipeded crystal, along its long direction which corresponds to an arbitrary crystallographic direction. In that case, the demagnetization factor was better characterized ($N \approx 3.2$ (cgs units)), and the value at the $M/H$ plateau did match with the $1/N$ value, confirming the ferromagnetic nature of the transition. 

\section{Magnetization curves}
\begin{figure}[h!]
\centerline{\includegraphics[width=7.0cm]{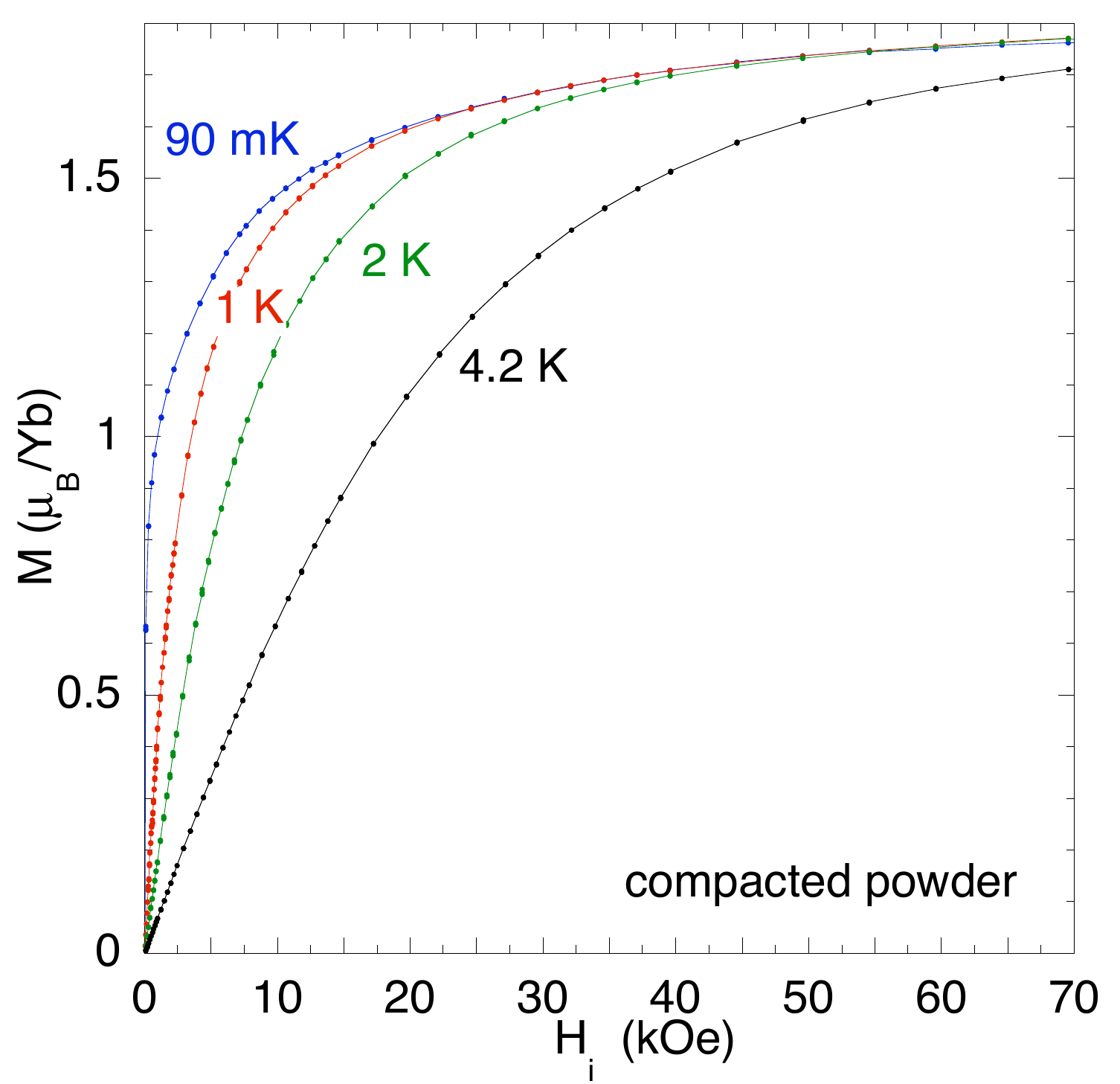}}
\caption{$M$ vs $H_i$ for the compacted powder for temperature between 90 mK and 4.2 K. The internal field $H_i=H-NM$ was calculated with $N=1.68$ (cgs units) from $M$ vs $T$ measurements.  }
\label{fig_pwd_MH}
\end{figure}

The magnetization curves up to 80 kOe and down to 90~mK were performed for the compacted powder and with the field aligned along the [100] and [110] directions for the single crystal. Below 2 K, the magnetization is almost saturated above 60 kOe and reaches about 1.75~$\mu_B/$Yb. 

Results for the compacted powder are shown in figure~\ref{fig_pwd_MH}, where the field has been corrected for demagnetizating effects. We assumed that the demagnetizing factor is equal to the inverse value of the $M/H$ plateau at low temperature in figure \ref{fig_MT_HT}, that is to say $N=1.68$ (cgs units) (thus slightly below the estimated value from the sample shape). 

\begin{figure}[h!]
\centerline{\includegraphics[width=7.0cm]{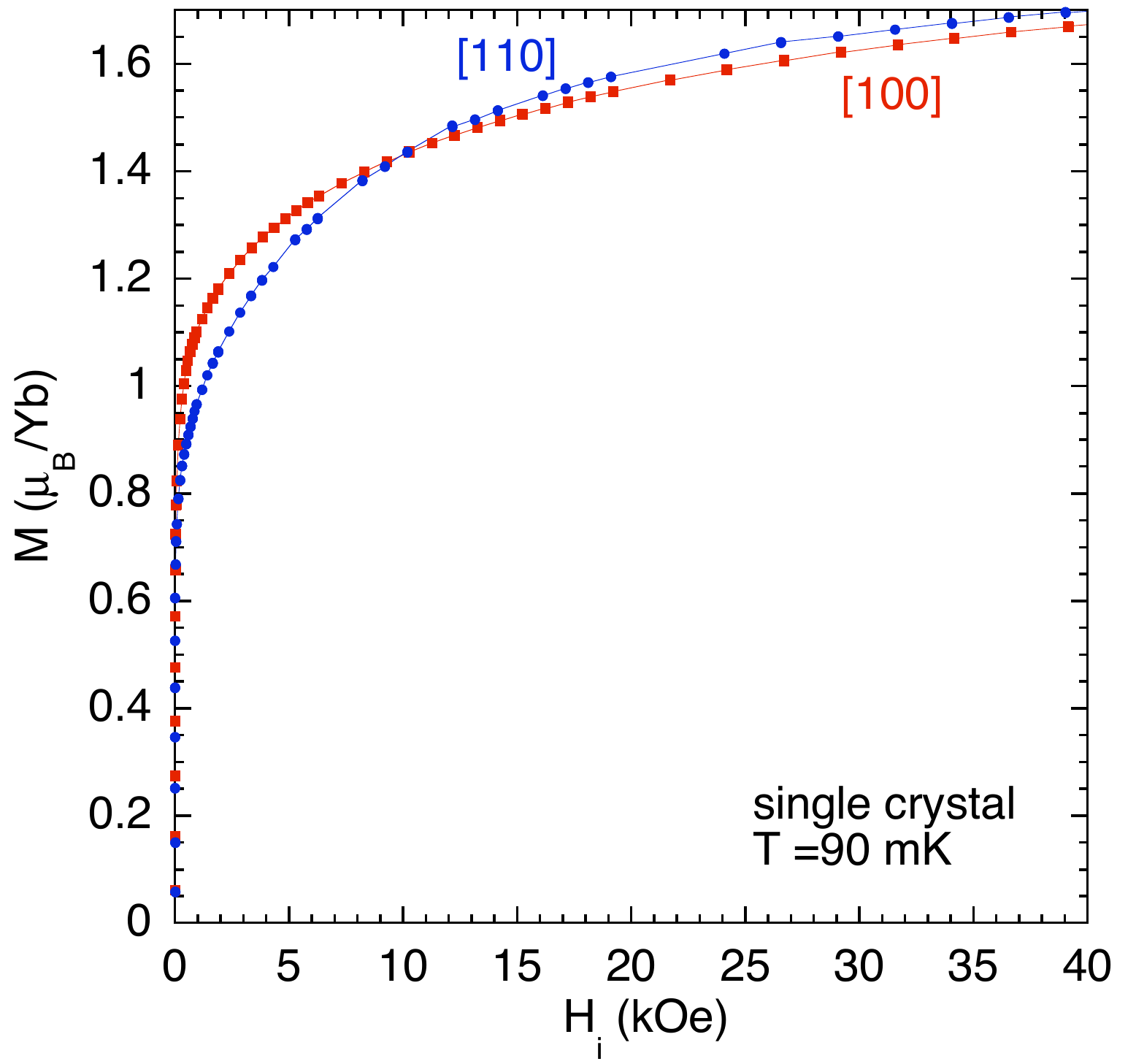}}
\caption{$M$ vs $H_i$ for the single crystal at 90 mK. The internal field $H_i=H-NM$ was calculated with $N_{[100]}=3.38$ and $N_{[110]}=3.8$ (cgs units) from $M$ vs $T$ measurements.  }
\label{fig_Xtal_MH}
\end{figure}

Above 2 K, no clear anisotropy is observed. Below 1 K, the magnetization curves start to separate significantly. In particular, in the low field region, the magnetization increases much faster when the field is applied along the [100] direction, as can be seen in figure~\ref{fig_Xtal_MH} (the demagnetization corrections were made using the same procedure as the powder). 

\section{Irreversibilities and Hysteresis}
\begin{figure}[h!]
\centerline{\includegraphics[width=7.0cm]{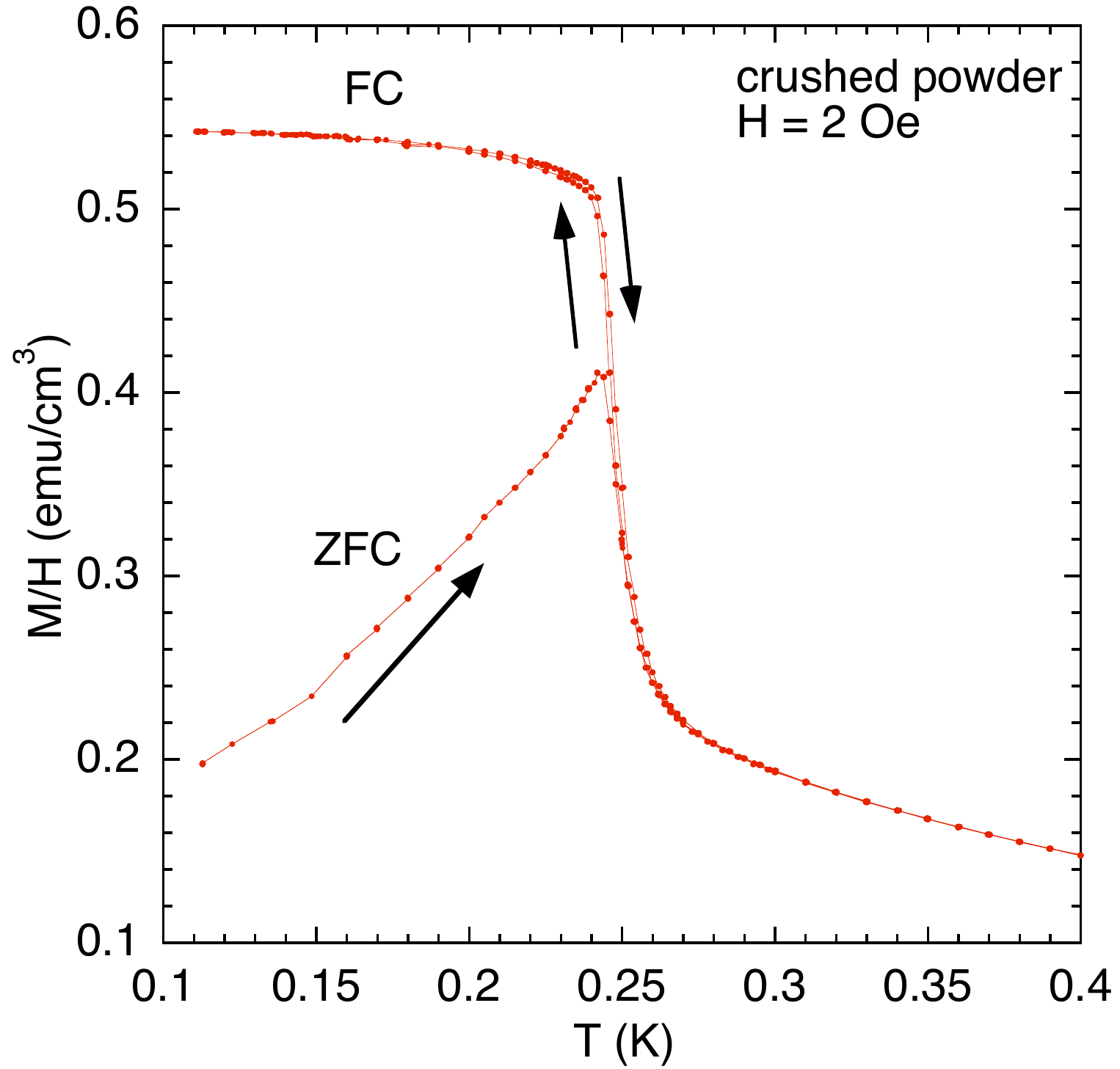}}
\caption{$M/H$ vs $T$ for the crushed powder for $H=2$ Oe measured in a ZFC-FC procedure. }
\label{fig_ZFC-FC}
\end{figure}

Zero Field Cooled - Field Cooled (ZFC-FC) measurements of the magnetization as a function of temperature exhibit an irreversibility (see figure \ref{fig_ZFC-FC}). This irreversibility is suppressed when the field is increased and disappears for fields larger than 500 Oe. This can be understood as the freezing of domains in the ferromagnetic state. (Note that all the measurements shown in the main article were measured in field cooled conditions.)

In addition, hysteresis loops were performed. A small hysteresis opens below the transition. In the powder samples, the hysteresis is about 20 Oe in width at 80 mK (giving a coercive field of 10 Oe) and closes at about 500~Oe. The hysteresis is 10 Oe in width at 200 mK and disappears at the transition. In the single crystal, with the field applied along [100], the hysteresis is smaller (less than 5 Oe width at 80 mK). 

These results are consistent with the ZFC-FC measurements. They imply that the domain pinning is quite weak and that the reversal of the magnetization in an applied field mainly occurs by a continuous rotation of the moments. 

\section{Determination of the spontaneous magnetization}
In a ferromagnetic LRO phase, the ferromagnetic ordered component can be associated with the spontaneous magnetization which is in turn deduced from magnetization curves $M$ vs $H$, provided that the field is applied in the direction of the ordered moment. 

In \YTO, the determination of this ordered component appears especially difficult, since the magnetic structure of the LRO phase is not known precisely. 

To address this question, we have analysed our $M$ vs $H$ magnetisation curves for the powders and for the single crystal measured along the [100] direction, and using a less comprehensive data along the [110] direction, as we did not perform precise measurements at very low fields. It is worth noting that the magnetization as a function of field deviates very quickly from the demagnetization line ($=H/N$), in agreement with the fast decrease of the $M/H$ plateau value observed in the $M$ vs $T$ curves when the field is increased (See figure 3 of the main article). This indicates that the value of the spontaneous moment is quite small. 

To make quantitative comparisons, we have plotted the magnetization as a function of the internal field $H_i$, supposing, as stated previously, that the demagnetizing factors $N$ equal the values of $H/M$ in very low field. 

The first conclusion is that the [100] direction seems to be the easy axis of magnetization, suggesting that that the moments are mainly oriented along the [100] direction. This result seems reasonable since it is approximately the obtained direction for the magnetic moments in Yb$_2$Sn$_2$O$_7$\cite{Yaouanc13}.

\begin{figure}[h!]
\centerline{\includegraphics[width=7.0cm]{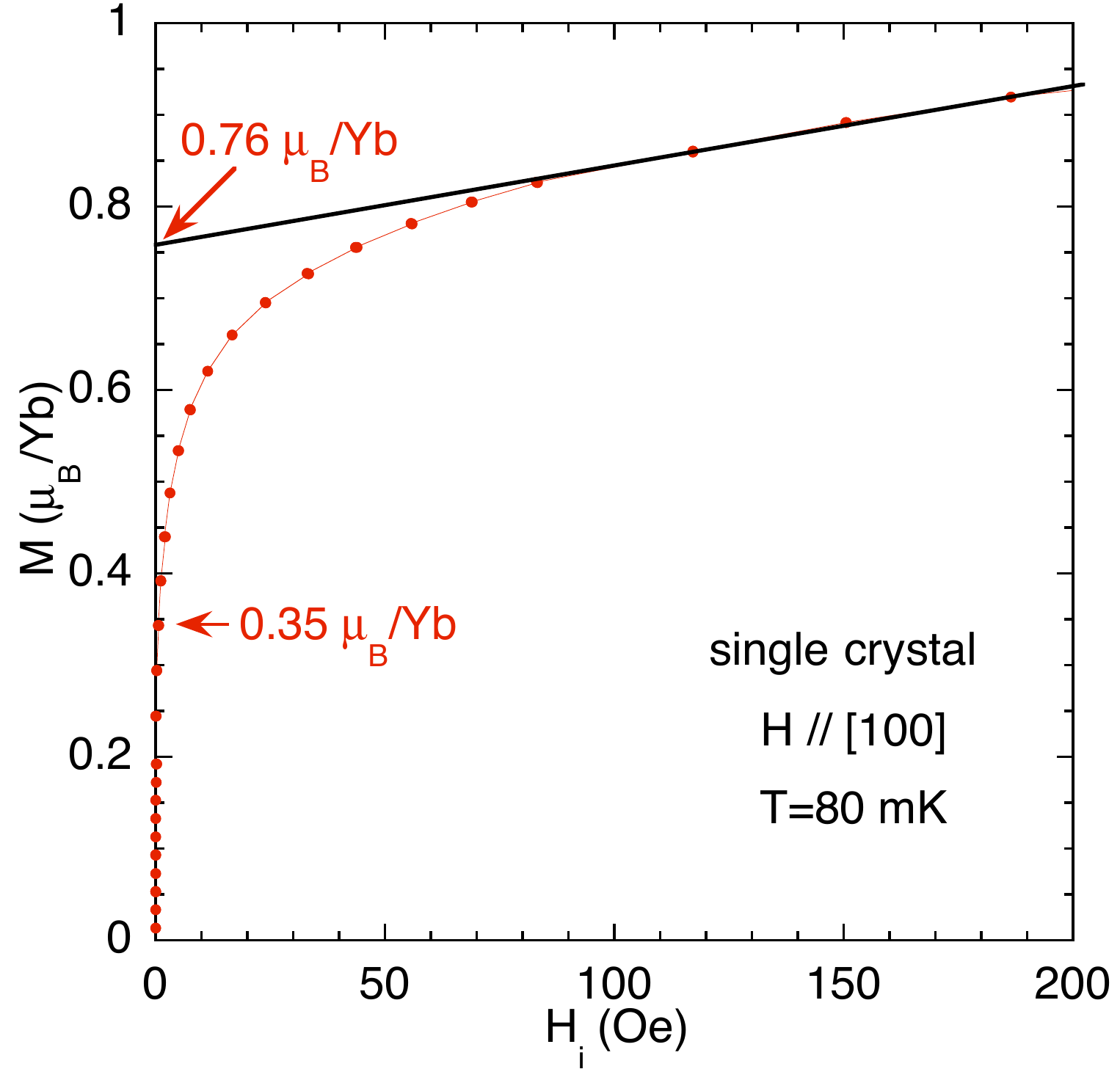}}
\caption{$M$ vs $H_i$ for the single crystal at $T=80$ mK and with $H$ applied along the [100] direction. $H_i$ was calculated with $N_{[100]}=3.38$ (cgs units). }
\label{fig_Ms}
\end{figure}

So, supposing that the ferromagnetic component is parallel to the [100] direction, we can estimate the spontaneous moment from the low temperature curve as shown in figure \ref{fig_Ms}. Strictly speaking, the spontaneous magnetization $M_s$ corresponds to the value at which the curve leaves the zero internal field. This would lead in the present case to a spontaneous moment of about 0.35~$\mu_B$/Yb. 
However, due to the curvature around zero internal field, an other criterion is usually considered: the spontaneous magnetization may be interpreted as the intercept of the slope of $M$ vs $H_i$ curve at larger field with the zero field axis. This has been done in figure \ref{fig_Ms} and gives a value of about 0.8 $\mu_B$/Yb. Nevertheless, here, there is no clear breakdown in the magnetization curve in any field range which would fix the field range to take into account for the extrapolation. This smooth curve shape may be due to the presence of a fluctuating component in the magnetization. 

The above analysis gives a ferromagnetic ordered moment in the range of 0.35 and 0.8 $\mu_B$/Yb at 80 mK in the single crystal, to be compared to the 1.75 $\mu_B$/Yb value of the magnetization in high field. The same kind of analysis in the powder gives a moment three times smaller, which is consistent with the hypothesis of an ordered moment along [100]. 
A possible temperature dependence of the spontaneous moment must be considered.
Even if it is the case that the transition is first-order, an increase in the ordered moment may be expected as the temperature is reduced further below $T_C$.
For the single crystal, our analysis was carried out at 80 mK which corresponds to $T_C/2$. The same procedure was followed at 110 mK and the results were found to be comparable.
For the powder sample, we performed the analysis between 80 (about $T_C/3$) and 200 mK, and no significant dependence of the spontaneous moment with temperature was observed.
These results suggest that the spontaneous moment will not increase significantly at lower temperature and confirm the first-order nature of the transition.

\end{document}